\documentclass[journal]{IEEEtran}

\ifCLASSINFOpdf
\else
   \usepackage[dvips]{graphicx}
\fi
\usepackage{url}

\usepackage{graphicx}
\usepackage[hyphenbreaks]{breakurl}
\usepackage{amsmath,amssymb,amsfonts}
\usepackage{subfigure}
\usepackage[colorlinks, citecolor = blue, linkcolor = blue]{hyperref}

\newtheorem{definition}{\bf Definition}

\begin{document}

\title{Multi-dimensional graph fractional Fourier transform and its application}

\author{Fang-Jia~Yan and Bing-Zhao~Li$^{*}$,~\IEEEmembership{Member,~IEEE}
\thanks{This work was supported in part by the National Natural Science Foundation of China under Grant 61671063.}
\thanks{Fangjia Yan is with the School of Mathematics and Statistics, Beijing Institute of Technology, Beijing 102488, China (e-mail: fangjia$\_$yan@bit.edu.cn).}
\thanks{Bingzhao Li is with the School of Mathematics and Statistics, Beijing Institute of Technology, Beijing 102488, China, and also with Beijing Key Laboratory on MCAACI, Beijing Institute of Technology, Beijing 102488, China (e-mail: li$\_$bingzhao@bit.edu.cn).}}

\markboth{Journal of \LaTeX\ Class Files, Vol. 14, No. 8, August 2015}
{Shell \MakeLowercase{\textit{et al.}}: Bare Demo of IEEEtran.cls for IEEE Journals}
\maketitle

\begin{abstract}
Many multi-dimensional signals appear in the real world, such as digital images and data that has spatial and temporal dimensions. How to show the spectrum of these multi-dimensional signals correctly is a key challenge in the field of graph signal processing. This paper investigates the novel transform for multi-dimensional signals defined on Cartesian product graph and studies several related properties. Our work includes: (i) defining the multi-dimensional graph fractional Fourier transform (MGFRFT) based on Laplacian matrix and adjacency matrix; (ii) exploring the advantages of MGFRFT in processing multi-dimensional signals in terms of spectrum and computational time; (iii) applying the proposed transform to data compression to highlight the utility and effectiveness of it.

%In this paper, we propose the multi-dimensional graph fractional Fourier transforms (MGFRFTs) using two basic graph signal processing methods i.e. based on Laplacian matrix and adjacency matrix. MGFRFTs are powerful tools for processing multi-dimensional signals. Compared with one-dimensional, it can avoid the multi-value of the spectrum.

%Then, the advantages of  two-dimensional transforms are extended to multi-dimensional graph fractional Fourier transforms. These new transforms have obvious advantages in processing multi-dimensional signals compared with traditional transforms on graph. Finally, we apply our new transforms to data compression.
\end{abstract}

\begin{IEEEkeywords}
Graph signal processing, graph Fourier transform, fractional Fourier transform, multi-dimensional signal processing.
\end{IEEEkeywords}

\IEEEpeerreviewmaketitle

\section{Introduction}
\label{intro}
\IEEEPARstart{U}{tilizing} the underlying structure in the data is the core of many signal processing tasks. In many cases, data resides on irregular domains. To describe the structure and interpret the complex relationships of network data sets in non-Euclidean spaces, some fundamental signal processing methods have been extended to graph signals \cite{ceci2020graph, gama2020graphs, morency2021graphon, 1, 2, 3}. Graphs become an important form of signal representation since they can be used to describe the geometry and topology of this kind of signal. In recent years, graph signal analysis has gained considerable attention, giving rise to the graph-based transforms such as graph Fourier transform (GFT) \cite{5,4}, graph filters \cite{Ramakrishna2020, 7}, sampling and recovery on graphs \cite{9, 8}. Judging from the existing research results, GFT mainly includes two basic signal processing methods i.e. based on Laplacian matrix and adjacency matrix.

Similar to the classical Fourier transform, it was shown that the GFT is inadequate for describing some applications or dealing with their underlying mathematical problems. As a result, the fractional Fourier transform (FRFT) on graph has been introduced to address the shortcoming of the GFT \cite{25, 26, 27, 28}. However, the spectral graph Fractional Fourier Transform (SGFRFT) and graph fractional Fourier transform (GFRFT) fail to show the correct spectrum of multi-dimensional (m-D) graph signals \cite{25, 27}. Because the eigenvalues of the graph Laplacian and the adjacency matrix act as the graph spectrum, in particular, the spectral functions are not well-defined, i.e., multi-valued when they have non-distinct eigenvalues \cite{2017Multi-}. Furthermore, given a graph, if the node number of it is large, the signal processing operations defined on this graph require more computational complexity. For example, the complexity of the SGFRFT \cite{27} using an eigendecomposition is $O(N^3)$ for a graph with N nodes. For better application to real data, it is urgent to find a new transform.

%This paper proposes two different multi-dimensional graph fractional Fourier transforms (MGFRFTs) for multi-dimensional graph signals, which solves the aforementioned problems related to traditional SGFRFT and GFRFT. First, MGFRFT can rearrange the 1-D spectrum obtained by the SGFRFT into the m-D frequency domain, and provide the m-D spectrum of the signal. Then, the Cartesian product graph can represent multi-domain data effectively \cite{Cartesian}. Whenever the underlying graph can be decomposed into two or more factor graphs with fewer nodes, the computational cost of these operations can be reduced significantly. To analyze the multi-dimensional graph signal and reduce computational complexity of it in graph fractional domain, we propose some new transforms. 

To solve the aforementioned problems related to traditional SGFRFT and GFRFT, we propose a transform that is suitable for multi-dimensional graph signal. The paper is organized as follows. Section \ref{Preliminary} overviews spectral graph theory and the definition of Cartesian product graphs. Section \ref{definitionmgfrft} defines the multi-dimensional graph fractional Fourier transform (MGFRFT). In addition, a performance analysis shows that compared with the traditional method, the spectrum accuracy and calculation efficiency of our new transform are significantly improved. Finally, we apply MGFRFT to data compression to clarify its application potentials in Section \ref{application}.

\section{Preliminaries}
\label{Preliminary}
\subsection{Spectral graph theory}
A weighted graph $\mathcal{G}=\{\mathcal{V}, \mathcal{E}, W\}$ consists of a finite set of vertices $\mathcal{V}=\{v_0, \cdots, v_{N-1}\}$, where $N=|\mathcal{V}|$ is the number of nodes, a set of edges $\mathcal{E}=\{(i,j)|i,j\in\mathcal{V},j\sim i\}\subseteq\mathcal{V}\times\mathcal{V}$, and a weighted adjacency matrix $W$. We assume that the graph is undirected with positive edge weights. The non-normalized graph Laplacian is a symmetric difference operator $\mathcal{L}=D-W$ \cite{13}, where $D$ is a diagonal degree matrix of $\mathcal{G}$. Let $\{\chi_0, \chi_1, \cdots, \chi_{N-1}\}$ be the set of orthonormal eigenvectors. 
Suppose that the corresponding Laplacian eigenvalues are sorted as $0=\lambda_0<\lambda_1\leq\lambda_2\leq\cdots\leq\lambda_{N-1}:=\lambda_{max}$.
We have $\mathcal{L}=\mathbf{\chi}\mathbf{\Lambda} \mathbf{\chi}^H$, where
	$\mathbf{\chi} = 
	[\chi_0, \chi_1, \cdots, \chi_{N-1}]$,
and the diagonal matrix is $\mathbf{\Lambda}=diag([\lambda_0, \lambda_1, \cdots, \lambda_{N-1}])$. The superscript $H$ represents the Hermitian transpose operation.

The fractional order is introduced to the Laplacian operator \cite{27}. The graph fractional Laplacian operator $\mathcal{L}_{\alpha}$ is defined by $\mathcal{L}_{\alpha}=\mathbf{\kappa}R\mathbf{\kappa}^{H}$, where $\alpha$ is the fractional order, $0<\alpha\leq1$,
$\ell=0,1,\cdots,N-1$ \cite{27}.
Note that $\mathbf{\kappa} = \begin{bmatrix}
		\kappa_0,& \kappa_1,& \cdots,& \kappa_{N-1}
	\end{bmatrix}={\chi}^\alpha$, 
and $R=\text{diag}({\begin{bmatrix}
			r_0,& r_1,& \cdots,& r_{N-1}
	\end{bmatrix}})
	=\mathbf{\Lambda}^\alpha$,
so that $r_\ell=\lambda_\ell^\alpha$. In the follow-up part of this paper, computing the $\alpha$ power of a matrix always uses matrix power function.

%The graph signal $f$ is defined as binding a scalar value to each node through the function $f:\mathcal{V}\rightarrow\mathbb{R}$. The SGFRFT of any signal $f$ building on the graph $\mathcal{G}$ is defined by \cite{27}:
%\begin{equation}
%	\begin{split}
%		\widehat{f}_{\alpha}(\ell)=\langle f,\kappa_{\ell}\rangle=\sum^{N}_{n=1}f(n)\kappa^{*}_{\ell}(n),  \ell=0,1,\cdots,N-1,
%	\end{split}
%\end{equation}
%when $\alpha=1$, the SGFRFT degenerates into standard GFT.

%The inverse SGFRFT is given by
%\begin{equation}\label{IGFRFT}
%	\begin{split}
%		f(n)=\sum^{N-1}_{\ell=0}\widehat{f}_{\alpha}(\ell)\kappa_{\ell}(n), n=0,1,\cdots,N-1.
%	\end{split}
%\end{equation}

\subsection{Cartesian product graph}\label{Cartesian}
Cartesian product graph is a type of graph multiplication \cite{Cartesian}. Consider two graphs $\mathcal{G}_1=\{\mathcal{V}_1, \mathcal{E}_1, W_1\}$ and $\mathcal{G}_2=\{\mathcal{V}_2, \mathcal{E}_2, W_2\}$. The graphs $\mathcal{G}_1$, $\mathcal{G}_2$ are factor graphs of the Cartesian product $\mathcal{G}_1\square\mathcal{G}_2$. $\mathcal{G}_1\square\mathcal{G}_2$ to be the graph with vertex set $\mathcal{V}_1\times\mathcal{V}_2$ and $\mathcal{V}_1=\{0, \cdots, N_1-1\}, \mathcal{V}_2=\{0, \cdots, N_2-1\}$. 

 The adjacency matrix $W_1\oplus W_2$ and Laplacian matrix $\mathcal{L}_1\oplus\mathcal{L}_2$ of $\mathcal{G}_1\square\mathcal{G}_2$ can be expressed by those of its factor graphs. 
\begin{equation}\label{W1W2}
	W_1\oplus W_2=W_1\otimes I_{N_2}+I_{N_1}\otimes W_2,
\end{equation}
\begin{equation}
	\mathcal{L}_1\oplus\mathcal{L}_2=I_{N_2}\otimes\mathcal{L}_1+\mathcal{L}_2\otimes I_{N_1},
\end{equation}
where $I_n$ is the identity matrix of size $n$ \cite{2020Learning}.

\begin{figure}[htbp]
	\centering
	\includegraphics[width=\linewidth]{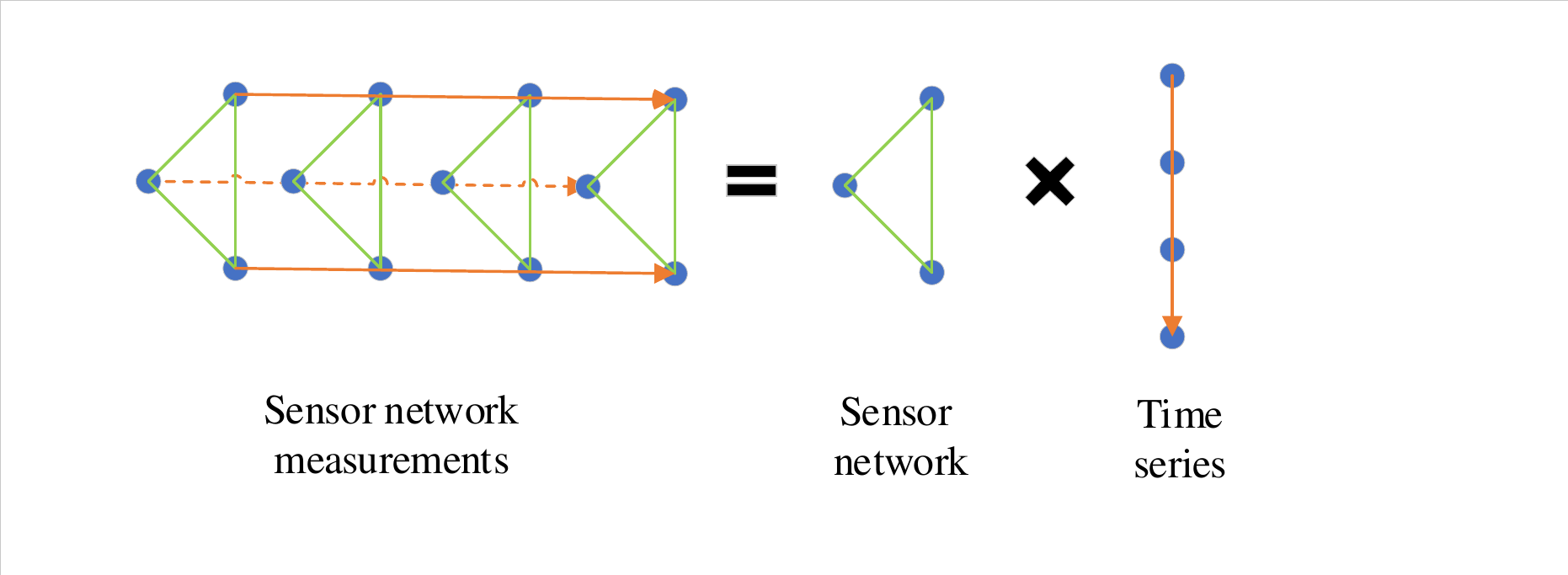}
	\caption{Measurements of a sensor network.}
	\label{product graph}
\end{figure}

We give an example of Cartesian product graph using in graph signal processing. In Fig. \ref{product graph}, the sensor network measurements are decomposed as the Cartesian product of sensor network and time series.

\begin{figure*}[t]
	\centering
	\subfigure[Original Signal]{
		\label{grid}
		\includegraphics[width=0.3\linewidth]{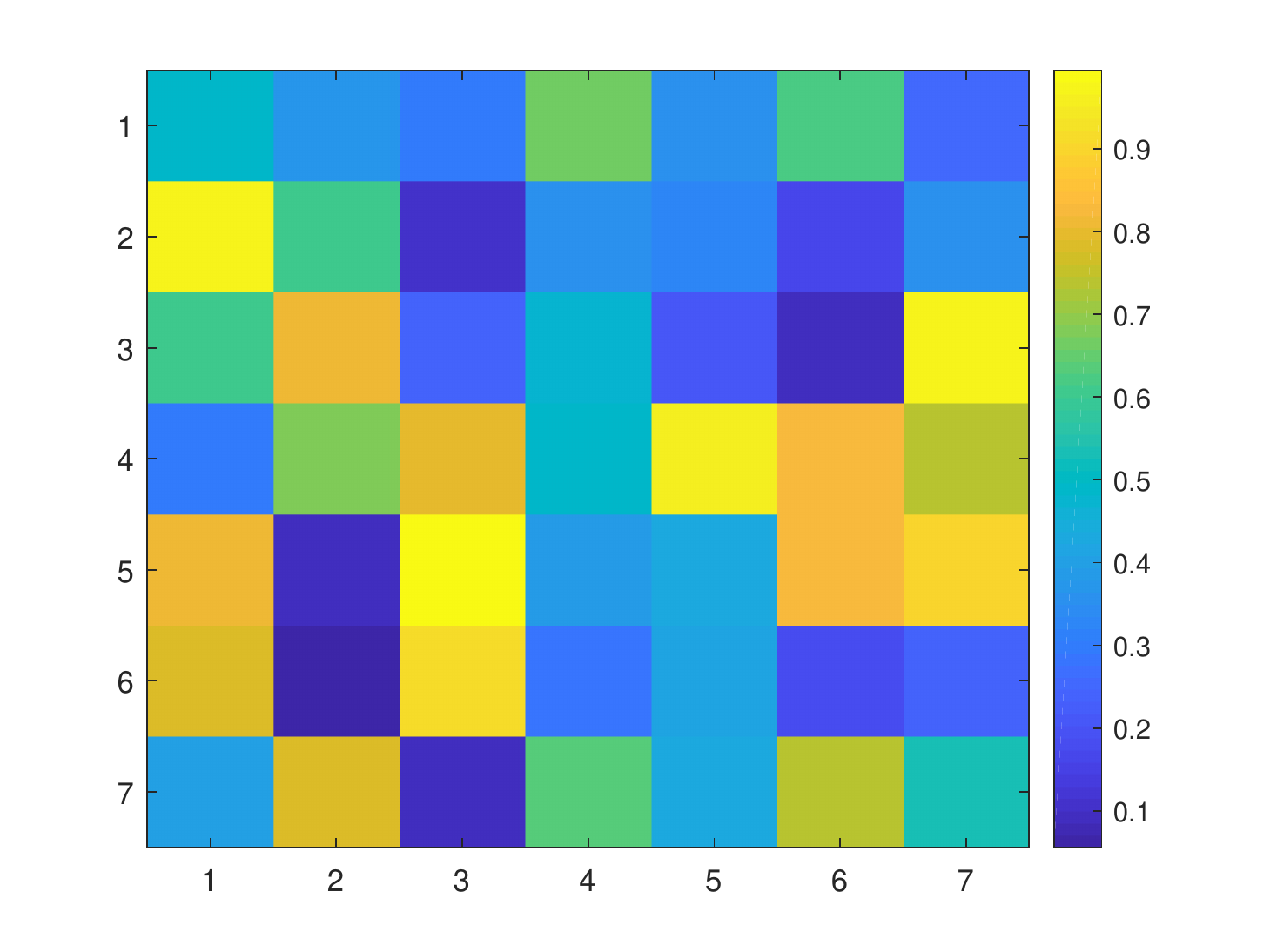}}
	\subfigure[Frequencies ]{
		\label{1dfre}
		\includegraphics[width=0.3\linewidth]{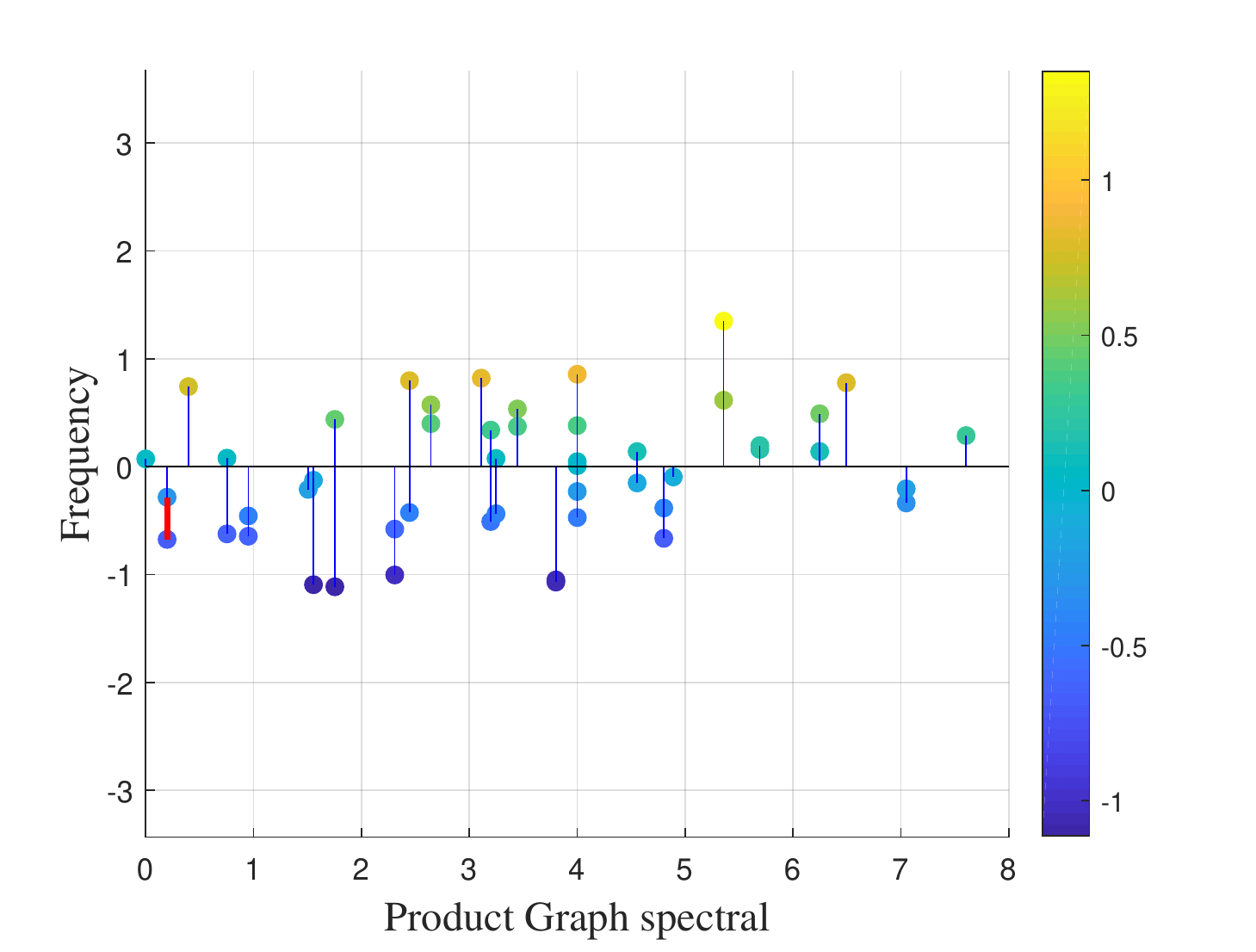}}
	\subfigure[2-D Frequencies]{
		\label{2dfre}
		\includegraphics[width=0.3\linewidth]{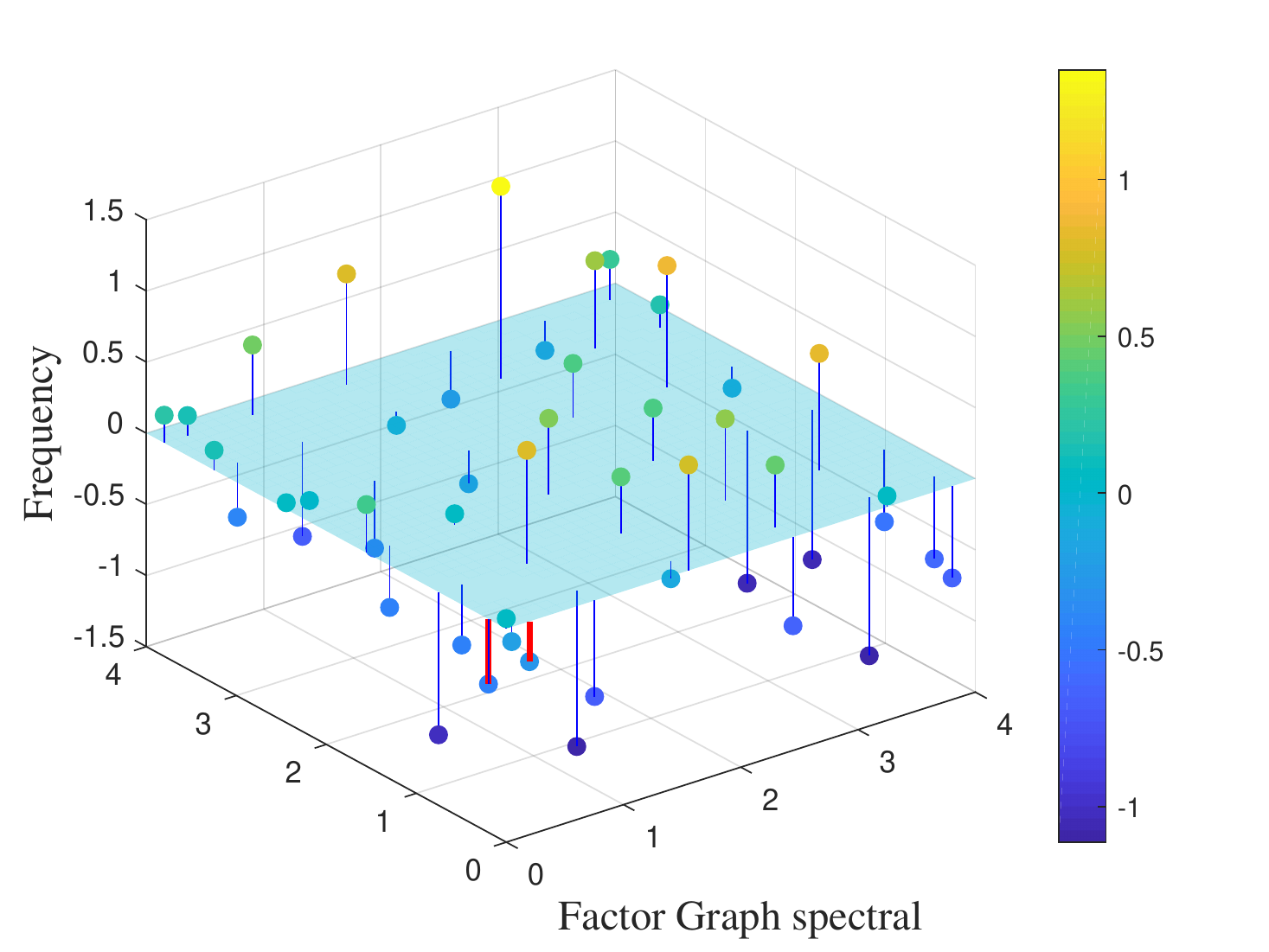}}
	\caption{2-D graph signal and its spectrum.}
	\label{multivalue}
\end{figure*}

\section{Multi-dimensional graph fractional Fourier transform}\label{definitionmgfrft}
In some cases, the spectrum is multi-valued in graph fractional domain. In order to show the spectrum more correctly, the definitions of multi-dimensional graph fractional Fourier transform (MGFRFT) based on Laplacian matrix and adjacency matrix are proposed in this section. The advantages of MGFRFT is also explored.

\subsection{Laplacian-based multi-dimensional graph fractional Fourier transform}\label{1L-MGFRFT}
Given a natural number $m$, consider $m$ graphs $\mathcal{G}_i=\{\mathcal{V}_i, \mathcal{E}_i, W_i\}$ with vertex set $\mathcal{V}_i$ and $\mathcal{V}_i=\{0, \cdots, N_i-1\}$, $i={1,2,\dots,m}$. $\mathcal{L}_i$ is the Laplacian matrix of $\mathcal{G}_i$ and its eigendecomposition is $\mathcal{L}_i=\mathbf{\chi}_i\mathbf{\Lambda}_i \mathbf{\chi}_i^H$. The corresponding graph fractional Laplacian operator is 
\begin{equation}
	\mathcal{L}_{\alpha}^{(i)}=\mathbf{\kappa}^{(i)}R^{(i)}(\mathbf{\kappa}^{(i)})^{H},
\end{equation}
where 
\begin{equation}
\mathbf{\kappa}^{(i)} = \begin{bmatrix}
		\kappa_0^{(i)},& \kappa_1^{(i)},& \cdots,& \kappa_{N_i-1}^{(i)}
	\end{bmatrix}={\chi}_i^\alpha,
\end{equation}
and 
\begin{equation}
R^{(i)}=\text{diag}({\begin{bmatrix}
			r_0^{(i)},& r_1^{(i)},& \cdots,& r_{N_i-1}^{(i)}
	\end{bmatrix}})
	=\mathbf{\Lambda}_i^\alpha.
\end{equation}

These $m$ graphs form a Cartesian product graph $\mathcal{G}_1\square\cdots\square\mathcal{G}_m$. Because the Cartesian product graph
\begin{equation}
	\begin{aligned}
		\mathcal{G}_1\square\cdots\square\mathcal{G}_m=((\cdots(\mathcal{G}_1\square\mathcal{G}_2)\cdots)\square\mathcal{G}_m),
	\end{aligned}
\end{equation}
we have the following eigendecomposition of the fractional Laplacian operator of  $\mathcal{G}_1\square\cdots\square\mathcal{G}_m$:

%Therefore, for $\mathcal{G}_1\square\cdots\square\mathcal{G}_m$, we have the following eigendecomposition of $\mathcal{L}_{\alpha}^{(1)}\oplus\cdots\oplus\mathcal{L}_{\alpha}^{(m)}$ in graph fractional domain：

\begin{equation}\label{fLeigendecomposition}
	\begin{aligned}
		&(\mathcal{L}_{\alpha}^{(1)}\oplus\cdots\oplus\mathcal{L}_{\alpha}^{(m)})\times\\
		&\begin{pmatrix}
			\kappa_{\ell_1}^{(1)}(0)\kappa_{\ell_2}^{(2)}(0)\dots\kappa_{\ell_m}^{(m)}(0)\\
			\kappa_{\ell_1}^{(1)}(0)\kappa_{\ell_2}^{(2)}(0)\dots\kappa_{\ell_m}^{(m)}(1)\\
			\vdots\\
			\kappa_{\ell_1}^{(1)}(N_1-1)\kappa_{\ell_2}^{(2)}(N_2-1) \dots\kappa_{\ell_m}^{(m)}(N_m-1)
		\end{pmatrix}\\
		={}&(r_{\ell_1}^{(1)}+\cdots+ r_{\ell_m}^{(m)})\times\\ 
		&\begin{pmatrix}
			\kappa_{\ell_1}^{(1)}(0)\kappa_{\ell_2}^{(2)}(0)\dots\kappa_{\ell_m}^{(m)}(0)\\
			\kappa_{\ell_1}^{(1)}(0)\kappa_{\ell_2}^{(2)}(0)\dots\kappa_{\ell_m}^{(m)}(1)\\
			\vdots\\
			\kappa_{\ell_1}^{(1)}(N_1-1)\kappa_{\ell_2}^{(2)}(N_2-1) \dots\kappa_{\ell_m}^{(m)}(N_m-1)
		\end{pmatrix}.
	\end{aligned}
\end{equation}

Define a signal on a Cartesian product graph as a "multi-dimensional signal". Using the eigenvalues and eigenfunctions of $\mathcal{L}_{\alpha}^{(1)}\oplus\cdots\oplus\mathcal{L}_{\alpha}^{(m)}$ in \eqref{fLeigendecomposition}, we propose the Laplacian-based multi-dimensional graph fractional Fourier transform (L-MGFRFT).
\begin{definition}
	The L-MGFRFT of a m-dimensional signal $f$ on a Cartesian product graph $\mathcal{G}_1\square\cdots\square\mathcal{G}_m$ is 
	\begin{equation}\label{L-MGFRFT}
		\begin{aligned}
			&\widehat{f}_{\alpha}(\ell_1+\cdots+\ell_m)\\
			={}&\sum^{N_1}_{n_1=1}\sum^{N_2}_{n_2=1}\dots\sum^{N_m}_{n_m=1}f(n_1,\dots,n_m)(\kappa^{(1)}_{\ell_1}(n_1))^*\cdots(\kappa^{(m)}_{\ell_m}(n_m))^*,
		\end{aligned}
	\end{equation}
	for $\ell_i=0,1,\cdots,N_i-1$. And its inverse (IL-MGFRFT) is given by 
	\begin{equation}
		\begin{aligned}
			&f(n_1,\dots,n_m)\\
			={}&\sum^{N_1}_{\ell_1=1}\sum^{N_2}_{\ell_2=1}\dots\sum^{N_m}_{\ell_m=1}\widehat{f}_{\alpha}(\ell_1+\cdots+\ell_m)\kappa^{(1)}_{\ell_1}(n_1)\cdots\kappa^{(m)}_{\ell_m}(n_m),
		\end{aligned}
	\end{equation}
	for $n_i=0,1,\cdots,N_i-1$.
\end{definition}

\subsection{Adjacency-based multi-dimensional graph fractional Fourier transform}
In addition to Laplacian-based method, there is another important method related to adjacency matrix in graph signal processing. Using the same Cartesian product graph as in Section \ref{1L-MGFRFT}, the eigendecomposition of the adjacency operator $W_i$ of graph $\mathcal{G}_i$ is $W_i=\mathbf{V_iJ_{W_i}V_i^{-1}}$. The eigendecomposition of adjacency matrix $W_1\oplus\cdots\oplus W_m$ is 
\begin{equation}
	\begin{aligned}
		&W_1\oplus\cdots\oplus W_m\\
		=&\mathbf{(V_1\otimes\cdots\otimes V_m)}\times\\
		&\mathbf{(\sum_{j=1}^m I_{N_1\dots N_{j-1}}\otimes J_{W_j}\otimes I_{N_{j+1}\dots N_{m}})}\times\\
		&\mathbf{(V_1\otimes\cdots\otimes V_m)^{-1}}.
	\end{aligned}
\end{equation}
Using the above eigendecomposition, we define adjacency-based multi-dimensional graph fractional Fourier transform (A-MGFRFT).

\begin{definition} The A-MGFRFT of a m-dimensional signal $\mathbf{f}$ on a Cartesian product graph $\mathcal{G}_1\square\cdots\square\mathcal{G}_m$ is
	\begin{equation}
		\begin{aligned}
			\mathbf{\hat{f}_\alpha}={}& \mathbf{((V_1\otimes\cdots\otimes V_m)^{-1})^\alpha f}.\\
			%={}& \mathbf{(V_1^{-1}\otimes\cdots\otimes V_m^{-1})^\alpha f}.
		\end{aligned}
	\end{equation}
	Its inverse (IA-MGFRFT) is given by 
	\begin{equation}
		\mathbf{f}=	\mathbf{(V_1\otimes\cdots\otimes V_m)^\alpha}\mathbf{\hat{f}_\alpha}.
	\end{equation}
\end{definition}

Though the spectrum obtained by these two MGFRFTs is usually different for the same graph signal $f$, the process of signal analysis is similar. Therefore, we mainly focus the Laplacian method in this paper. As for the adjacency method, we only give the definition.

\subsection{Advantage analysis}
For a signal on Cartesian product graph, the proposed MGFRFTs provide the following advantages over the traditional GFRFT and SGFRFT. We use the two-dimensional transform based on Laplacian matrix as an example to illustrate how the advantages are reflected in the graph signal processing.

As the spectrum of the multi-dimensional graph signal obtained by 1-D transforms is often multi-valued at a specific frequency and the GFRFT and SGFRFT can only show 1-d spectrum. These two traditional methods are not suitable for multi-dimensional graph signals. In Fig \ref{multivalue}, with $\alpha=0.9$, we visually display the shortcoming of the SGFRFT and the advantage of L-MGFRFT. Fig \ref{grid} is a random 2-D graph signal on the grid graph. In Fig \ref{1dfre}, we see that at certain frequencies, the spectrum is double-valued. We use red line to give an example of double-valued spectrum. In Fig \ref{2dfre}, all the spectrum are single-valued, and we use the same red color to show that the double-valued spectrum in Fig \ref{1dfre} can be identified and separated by our new transform. MGFRFT can rearrange the 1-D spectrum obtained by the SGFRFT into the m-D frequency domain, and provide the m-D spectrum of the signal. Therefore, MGFRFT can solve the multi-valuedness of the spectral functions.

Another advantage that can not be ignored is time consumption. The Cartesian product graph can represent multi-domain data effectively \cite{Cartesian}. Whenever the underlying graph can be decomposed into two or more factor graphs with fewer nodes, the computational cost of these operations can be reduced significantly. Considering a Cartesian product graph $\mathcal{G}_1\square\mathcal{G}_2$, the MGFRFT costs $O(N_1^2N_2+N_1N_2^2)$ time, whereas the traditional SGFRFT costs $O(N_1^2N_2^2)$ on the same graph. When we use factor graphs $\mathcal{G}_1$ and $\mathcal{G}_2$ instead of the original graph to calculate the eigendecomposition of the graph fractional Laplacian, it also takes less time.

\begin{figure}[t]
	\centering
	\includegraphics[width=\linewidth]{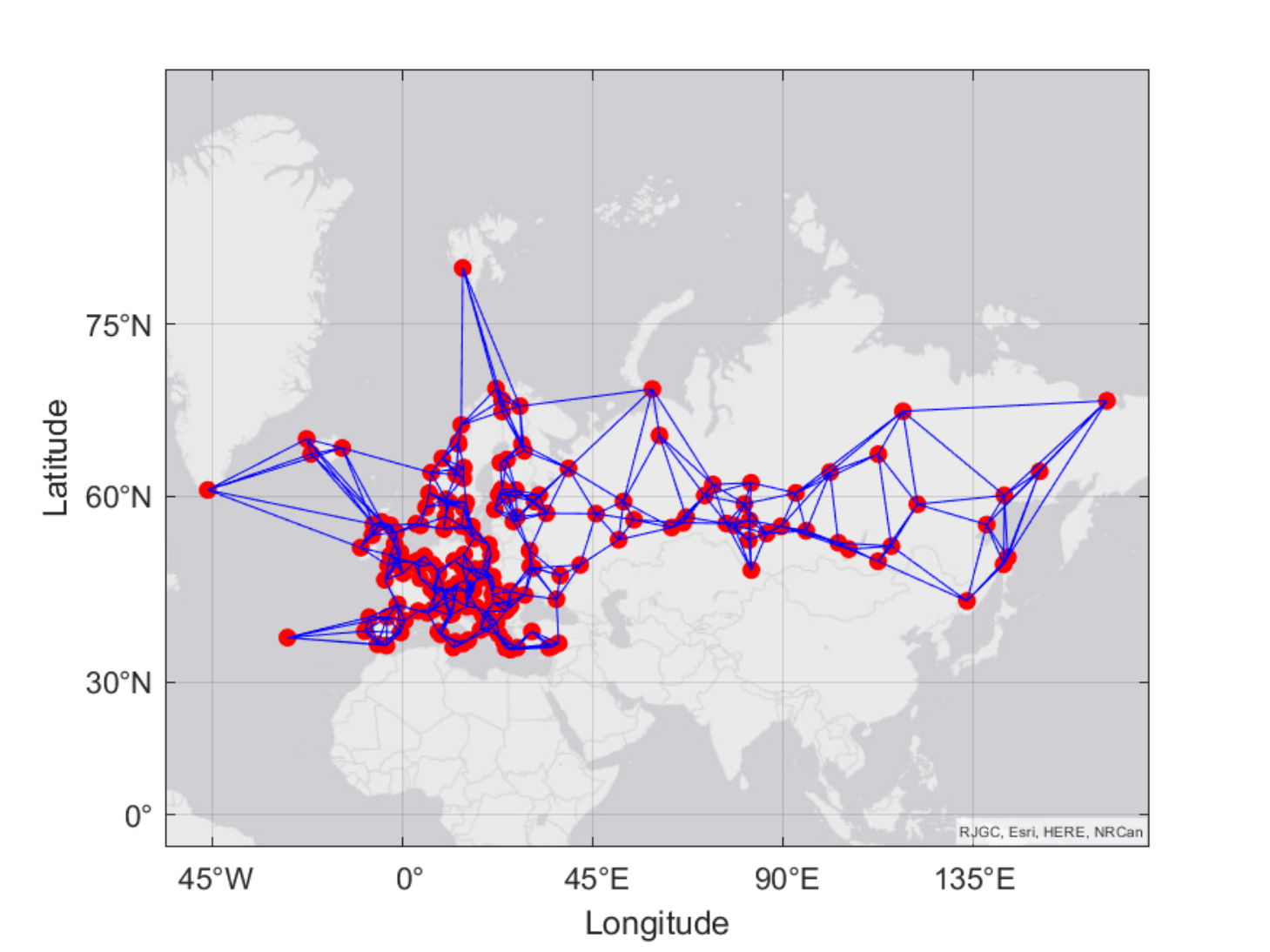}
	\caption{The $k$NN graph $S$ for weather observation station networks.}
	\label{fig:distance}
\end{figure}

\begin{figure*}[t]
	\centering
	\subfigure[Original Signal]{\includegraphics[width=0.3\linewidth]{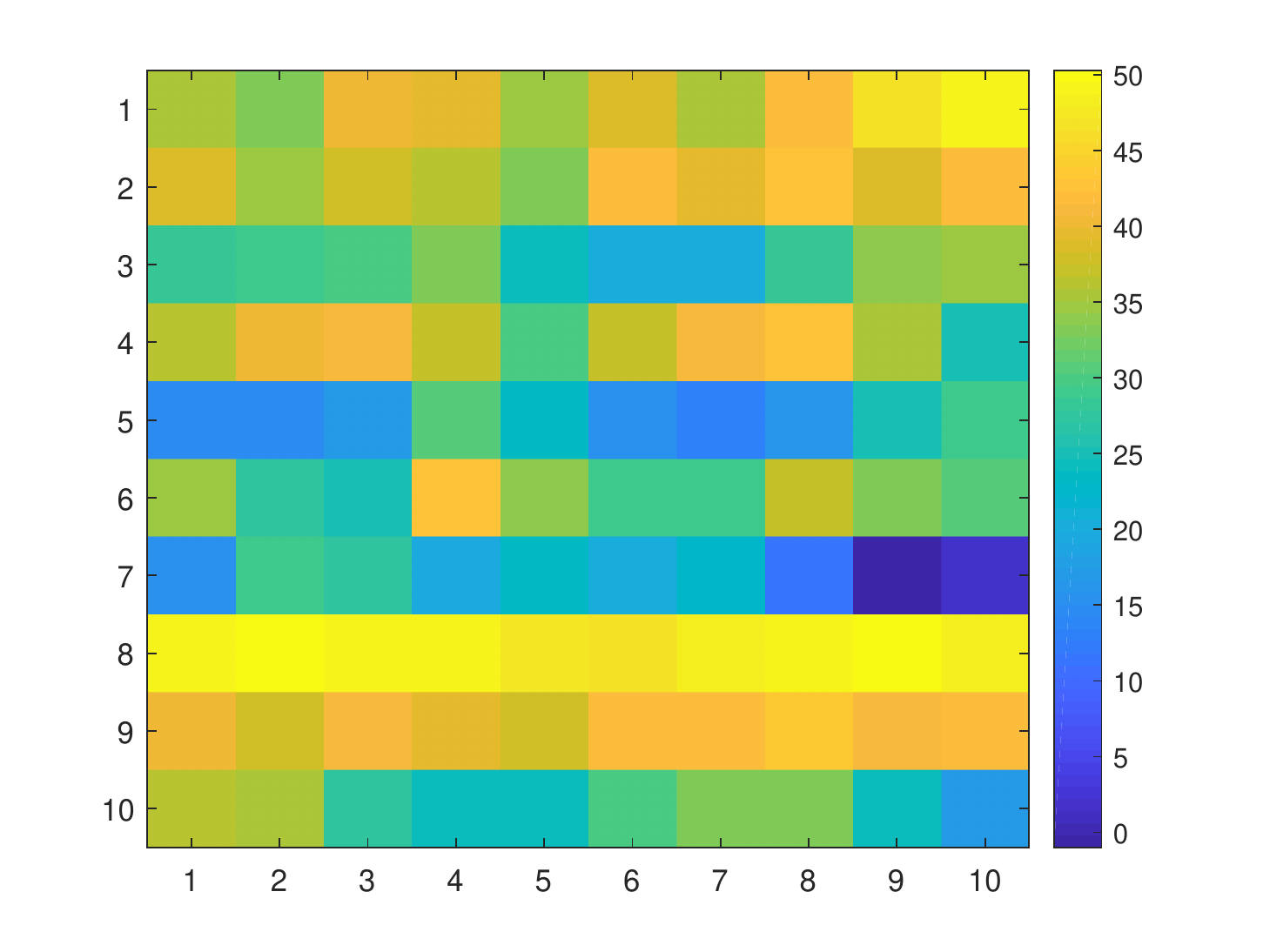}}
	\subfigure[MGFRFT Frequencies ]{\includegraphics[width=0.3\linewidth]{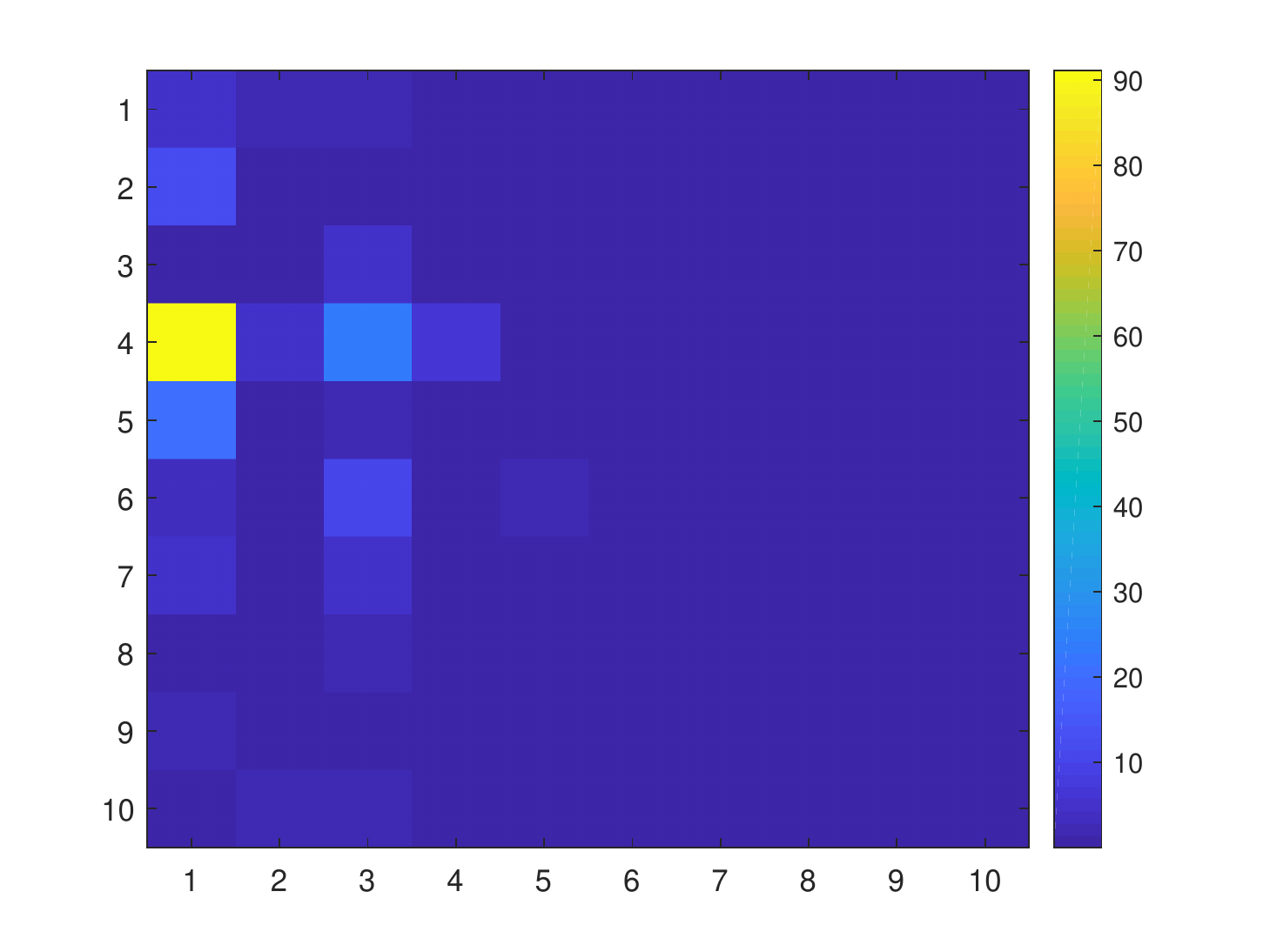}}
	\subfigure[Compressed Signal]{\includegraphics[width=0.3\linewidth]{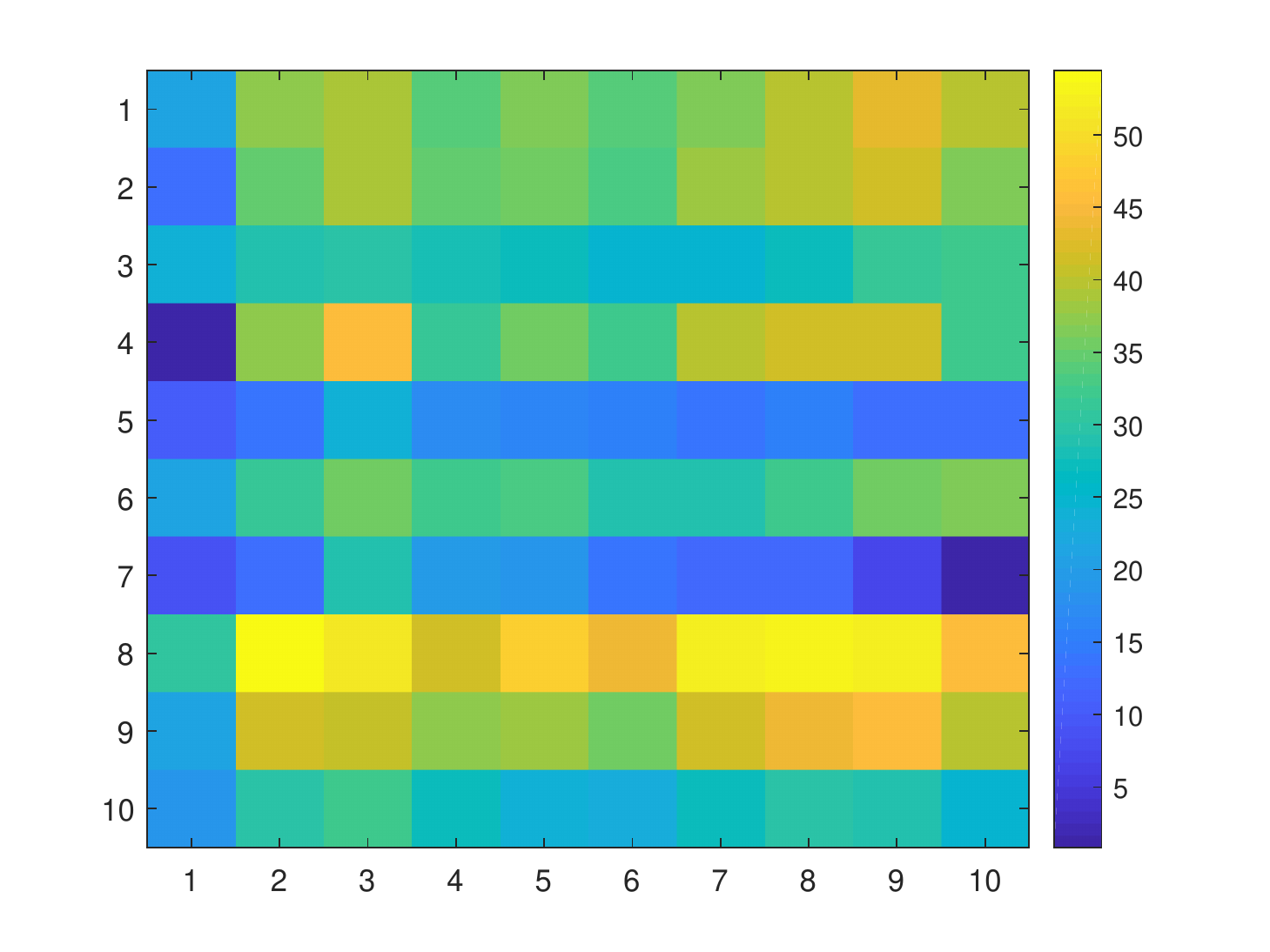}}
	\caption{An illustration of the compression process.}
	\label{fig:illustrate}
\end{figure*}

\begin{table*}
	\centering
	\caption{RE and PSNR for different compression parameters.}
	\begin{tabular}{|c|c|c|c|c|c|c|c|}
		\hline
		$\alpha = 0.95$ & $\gamma = 0.02$ & $\gamma = 0.05$ & $\gamma = 0.10$ & $\gamma = 0.15$ & $\gamma = 0.20$&$\gamma = 0.25$&$\gamma = 0.30$\\
		\hline
		RE(\%) & 10.49 & 8.56 & 7.00& 5.97& 5.18& 4.52& 3.96\\
		PSNR(dB) & 86.49& 88.25& 90.11& 91.67& 93.15& 94.46& 95.83\\
		\hline
		$\alpha = 0.85$ & $\gamma = 0.02$ & $\gamma = 0.05$ & $\gamma = 0.10$ & $\gamma = 0.15$ & $\gamma = 0.20$&$\gamma = 0.25$&$\gamma = 0.30$\\
		\hline
		RE(\%) & 18.52 & 15.65 & 12.99& 11.15& 9.69& 8.48& 7.43\\
		PSNR(dB) & 89.10& 86.80& 85.11& 85.59& 86.93& 88.20& 89.60\\
		\hline
	\end{tabular}
	\label{tab:RE}
\end{table*}

\section{Applications}\label{application}

To illustrate the effectiveness of the framework in MGFRFT, we consider an application to data compression. Using public real-world data from National Oceanic and Atmospheric Administration (NOAA)'s National Centers for Environmental Information (NCEI), we randomly select 200 weather observation stations and pick out temperature records in 2020. The data is available in \sloppy\burl{https://www.ncei.noaa.gov/data/global-summary-of-the-day/access/2020/}.

The temperature records naturally form a path graph with 365 nodes, denoted by $\mathcal{T}$. To construct a graph model for stations, we adopt $k$NN method as shown in \cite{knn}. For each pair of stations, we compute the spherical distance via latitudes and longitudes. Specifically, let $(\theta_i,\phi_i)$ and $(\theta_j,\phi_j)$ be the coordinates for two stations $i$ and $j$ (in radian measure). The spherical distance is defined by
\begin{equation}
	\begin{aligned}
			d_{ij}={}& \arccos\bigg(\cos(\theta_1)\cos(\theta_2)\cos(\phi_1-\phi_2)+\\
			&\sin(\theta_1)\sin(\theta_2)\bigg)
	\end{aligned}
\end{equation}

Note that we drop the earth radius $R=6,357$km in the computation of spherical distance since it will cause large precision error when we use the Gaussian kernel.

Then we search the $k$-nearest-neighbors for each station based on the spherical distance. The $k$NN graph $S$ is constructed by setting each node as a station and a pair of nodes is connected iff they are neighbors. Fig. \ref{fig:distance} shows the $k$NN graph where $k$ is set to be 5. 

The weighted matrix $W$ is defined by substituting the spherical distances into the Gaussian kernel. Let $i$ and $j$ be two nodes joined by an edge. The weight $W_{ij}$ is set to be

\begin{equation}
	W_{ij} = \frac{\exp(-\frac{d_{ij}^2}{\sigma^2})}{\sqrt{\sum_{k\sim i}\exp(-\frac{d_{ik}^2}{\sigma^2})}\sqrt{\sum_{k\sim j}\exp(-\frac{d_{jk}^2}{\sigma^2})}}
\end{equation} 
where $\sim$ denotes that two nodes are incident, and $\sigma^2$ is a preset parameter.

The temperature data thus can be regarded as a signal $f$ on the product graph $\mathcal{G}\square \mathcal{T}$ where each element $f(i,t)$ denotes the temperature at station $i$ on day $t$. The compression process is performed as follows. We compute the MGFRFT of $f$ using equation \eqref{L-MGFRFT}. Then we sort the coefficients according to their absolute values. Let $0<\gamma<1$ be the compression ratio. We retain $\gamma$ fraction of largest coefficients and set remaining coefficients to be zero. Finally we perform the inverse MGFRFT on the resulting coefficients. Fig. \ref{fig:illustrate} presents a $10\times 10$ block of the original temperature data and compressed temperature data.

Two quantities are used to measure the compression loss. Let $f_{com}$ be the compressed signal. The relative  error (RE) is defined by

\begin{equation}
	\text{RE} = \frac{\sum_{i\in \mathcal{G}, t\in \mathcal{T}}|f(i,t)-f_{com}(i,t)|}{\sum_{i\in \mathcal{G}, t\in \mathcal{T}}|f(i,t)|}.
\end{equation}  

The peak signal-to-noise ratio (PSNR) is defined by

\begin{equation}
	\begin{aligned}
		&\text{PSNR}\\
		={}& 10\log_{10}\left(\frac{\max_{i\in \mathcal{G}, t\in \mathcal{T}}|f_{com}(i,t)|^2}{\frac{1}{|\mathcal{G}||\mathcal{T}|}\sum_{i\in \mathcal{G}, t\in \mathcal{T}}|f(i,t)-f_{com}(i,t)|^2}\right).
	\end{aligned}
\end{equation}

We carry out various experiments with different compression ratio $\gamma$ and transform fractional orders $\alpha$. The result is shown in Tab. \ref{tab:RE}. The results show that even for extremely small $\gamma$, the compression only introduces a small relative error. Further, compared with the results shown in \cite{Moura2013IEEE}, the MGFRFT provides better transform basis and yields smaller error. Tuning the fractional order $\alpha$ is also helpful to get more plausible compression.

\section{Conclusions}

In this paper, an approach to multi-domain data is presented. First, two different definitions of MGFRFT are proposed. Then to make implementations of our new transforms suitable for processing multi-dimensional signals, we compare the frequency obtained by one-dimensional transform and two-dimensional transform visually. The results show that the spectrum of our MGFRFT is well-defined, which avoids the double-valued spectrum of one-dimensional transforms that sometimes occurs. Finally, the MGFRFT is applied to data compression, using real-world data. The experiment reveals that compression based on our new transform is more efficient.

\clearpage
\bibliographystyle{IEEEtran}
\bibliography{references}

\end{document}